\documentclass{aa}

\usepackage{graphicx}
\usepackage{txfonts}
\usepackage[english]{babel}
\usepackage{natbib}

\usepackage{lipsum}
\usepackage{color}
\usepackage{subcaption}
\usepackage{lscape}
\usepackage{placeins}
\usepackage{adjustbox}
\usepackage{longtable}
\usepackage{booktabs}
\usepackage{float}
\usepackage{rotating}
\usepackage{siunitx}

\usepackage{hyperref}

\begin{document}

   \title{Automating the detection of polarization angle rotations in blazars}

   \subtitle{Re-analysis of RoboPol data reveals 27 new rotations.}

\author{
Anastasia Glykopoulou \inst{\ref{AstroCrete2},\ref{AstroCrete}}\thanks{\href{mailto:glykopoulouanastasia@gmail.com}{glykopoulouanastasia@gmail.com}},
Ioannis Liodakis \inst{\ref{AstroCrete}},
Dmitry Blinov \inst{\ref{AstroCrete2},\ref{AstroCrete}}
}

\institute{
Department of Physics, University of Crete, GR-70013 Heraklion, Greece \label{AstroCrete2}
\and
Institute of Astrophysics, Foundation for Research and Technology-Hellas, Vasilika Vouton, GR-70013 Heraklion, Greece \label{AstroCrete}
}

\abstract
   {We present an automated pipeline for the detection of EVPA rotations in blazars, integrating correction of the 180$^\circ$ ambiguity, Bayesian Blocks segmentation, and statistical validation.
   Applied to RoboPol monitoring data, the method identified 48 rotations across 25 sources, including multiple events in RBPLJ2232+1143, RBPLJ1751+0939, RBPLJ1800+7828, and RBPLJ2253+1608.
   The rotations span amplitudes from $90.8^{\circ}$ to $359.7^{\circ}$, durations between 7.0 and 111.3 days, and rotation rates averaging {$5.0^{\circ}$/day}. Comparison with previous catalogs reveals systematic differences: Bayesian Blocks rotations are on average $\sim 10\%$ larger in amplitude, about twice as long in duration, and roughly two-thirds slower in angular velocity, reflecting systematic biases between adaptive binning and manual segmentation. In addition, we report 27 previously unreported rotations, including 11 from the final 2016--2017 season. A correlation analysis with contemporaneous Fermi--LAT $\gamma$-ray light curves shows that longer rotations tend to coincide with enhanced $\gamma$-ray activity, while rotation amplitude alone is not predictive of $\gamma$-ray brightness. Our pipeline minimizes subjective biases, expands the list of known EVPA rotations, and provides a
   reproducible framework for future multiwavelength studies of blazar jet dynamics and particle acceleration.}


\date{Received December 2, 2025; accepted May 18, 2026}

\maketitle
\nolinenumbers

\section{Introduction}

Blazars are a subclass of active galactic nuclei (AGN) whose powerful relativistic jets are oriented within a few degrees from the observer's line of sight on Earth \citep{Blandford2019,Hovatta2019}, producing bright and strong variability across the electromagnetic spectrum. The optical emission is
dominated by synchrotron radiation from the jet, and therefore is highly polarized. Optical polarization monitoring programs have revealed that the electric vector position angle (EVPA) of the jet often undergoes large and systematic rotations, which provide a unique probe of the magnetic field structure and particle acceleration processes in jets. Such rotations have been associated with shocks \citep[e.g.,][]{Marscher2008}, turbulence \citep[e.g.,][]{Marscher2014}, or geometrical changes in
the jet \citep{Abdo2010,Raiteri2017-II,Britzen2018}, and have been found to statistically coincide with enhanced $\gamma$-ray activity \citep{Blinov2018}, suggesting a physical connection between polarization variability and high-energy emission.

The first systematic study of EVPA rotations was carried out in the framework of the RoboPol program \citep{Angelakis2016,blinov2015,blinov2016b,blinov2016a,Blinov2018,Blinov_2019}, which provided dense, multi-epoch polarimetric monitoring of a large blazar sample \citep{Pavlidou2014}. These efforts established the first statistical catalogs of EVPA rotations, employing well-defined algorithmic procedures for event identification. While effective, these methods remained sensitive to measurement uncertainty and intrinsic variability, which could influence the outcome.
The highly stochastic nature of blazar jet variability \citep{Marscher2014,kiehlmann2017_roboPolEVPA} therefore warrants more robust identification techniques. More recently, long-term monitoring of $\gamma$-ray blazars has emphasized the importance of reproducible methodologies to quantify such correlations across large samples \citep{2023MNRAS.523.4504O,Savchenko_2024}.

Motivated by \citet{Savchenko_2024}, we developed a fully automated pipeline tailored to the RoboPol blazar sample, capable of resolving the $180^{\circ}$ ambiguity, segmenting time series with Bayesian Blocks, and applying rigorous statistical tests.

In this work we present the pipeline and its application to the RoboPol dataset, producing a reproducible catalog of rotation events, recovering previously overlooked cases, and extending the sample into the final RoboPol observing season 2016--2017 that had not been previously used to detect rotations. By comparing our results with earlier catalogs and examining correlations with contemporaneous $\gamma$-ray activity from Fermi--LAT \citep{Abdollahi_2023}, we aim to provide new insights into the physical mechanisms driving EVPA rotations and their role in blazar jet dynamics.

\section{Data and Methods}
\subsection{Data sample and observations}

The dataset used in this analysis originates from the RoboPol monitoring data described by \citet{Blinov2021}, which includes all sources of the main and control samples as well as individual sources observed with RoboPol over the duration of the program that were not part of the statistical analysis. Details on the monitoring program can be found in \cite{Pavlidou2014} \& \cite{Angelakis2016} , and the data analysis procedures in \cite{King2014,Panopoulou2015,Blinov2021}.

The monitoring campaign lasted from 2013--2017 and provides multi-epoch polarimetric measurements with dense temporal coverage, enabling the study of EVPA variability and long-term polarization behavior. In addition to the optical polarization data, we made use of the \textit{Fermi}{-LAT} light curve repository \citep{Abdollahi_2023} to examine the contemporaneous $\gamma$-ray activity of the sources and assess potential correlations with the optical polarization behavior.

\subsection{Pipeline Structure}

\subsubsection{Preprocessing}

The monitoring data were ingested from CSV files containing the polarization measurements of each source. Julian dates were converted to Modified Julian Dates (MJD) for convenience.
Time differences between consecutive observations were computed to facilitate the identification of observing gaps. To ensure statistical reliability, only measurements with a polarization degree significance of at least three ($\mathrm{PD}/\mathrm{err}_\mathrm{PD} \geq 3$) were retained. The resulting dataset thus contains chronologically ordered, high-quality observations with consistent time stamps, forming the basis for subsequent segmentation and EVPA analysis.

\subsubsection{EVPA $180^\circ$ Ambiguity}

EVPA measurements are defined modulo 180$^\circ$, which can introduce artificial discontinuities in time series analysis. To avoid spurious rotations, we implemented an error-weighted adjustment procedure that explicitly accounts for EVPA measurement uncertainties e.g.\citep{Blinov2018}. Gaps larger than 30 days between consecutive observations prevent reliable angle disambiguation; such gaps are therefore used as natural segment boundaries (see also the section on rotation detection below).

For each new data point, the measured angle was compared to the previously adjusted value, taking into account the combined uncertainty of the two measurements. If the difference was smaller than ninety degrees after weighting by the uncertainties, the angle was retained. Otherwise, integer multiples of 180$^\circ$ were added to the measured angle, and the candidate value that minimized the weighted difference with respect to the previous adjusted angle was selected.

\subsubsection{Bayesian Blocks Analysis}

For the segmentation of the EVPA time series we employed the Bayesian Blocks algorithm
\citep{2013ApJ...764..167S}, using the implementation provided in the \texttt{astropy.stats} package \citep{2022ApJ...935..167A}. The input consisted of the observation times (MJD), the adjusted EVPA values, and their measurement uncertainties. Since the EVPA data represent continuous measurements with Gaussian errors, we adopted the \texttt{fitness='measures'} option. The sensitivity to change points was controlled through the false-alarm probability parameter $p_{0}$, which we set to $0.001$. This configuration yields an adaptive segmentation of the EVPA time series into statistically homogeneous blocks, suitable for subsequent rotation and extrema analysis.

\subsubsection{Identification of Local Extrema and Rotation Detection Algorithm}

\label{sec:rotation_detection}

\begin{figure*}[!ht]
    \centering
    \includegraphics[width=\linewidth]{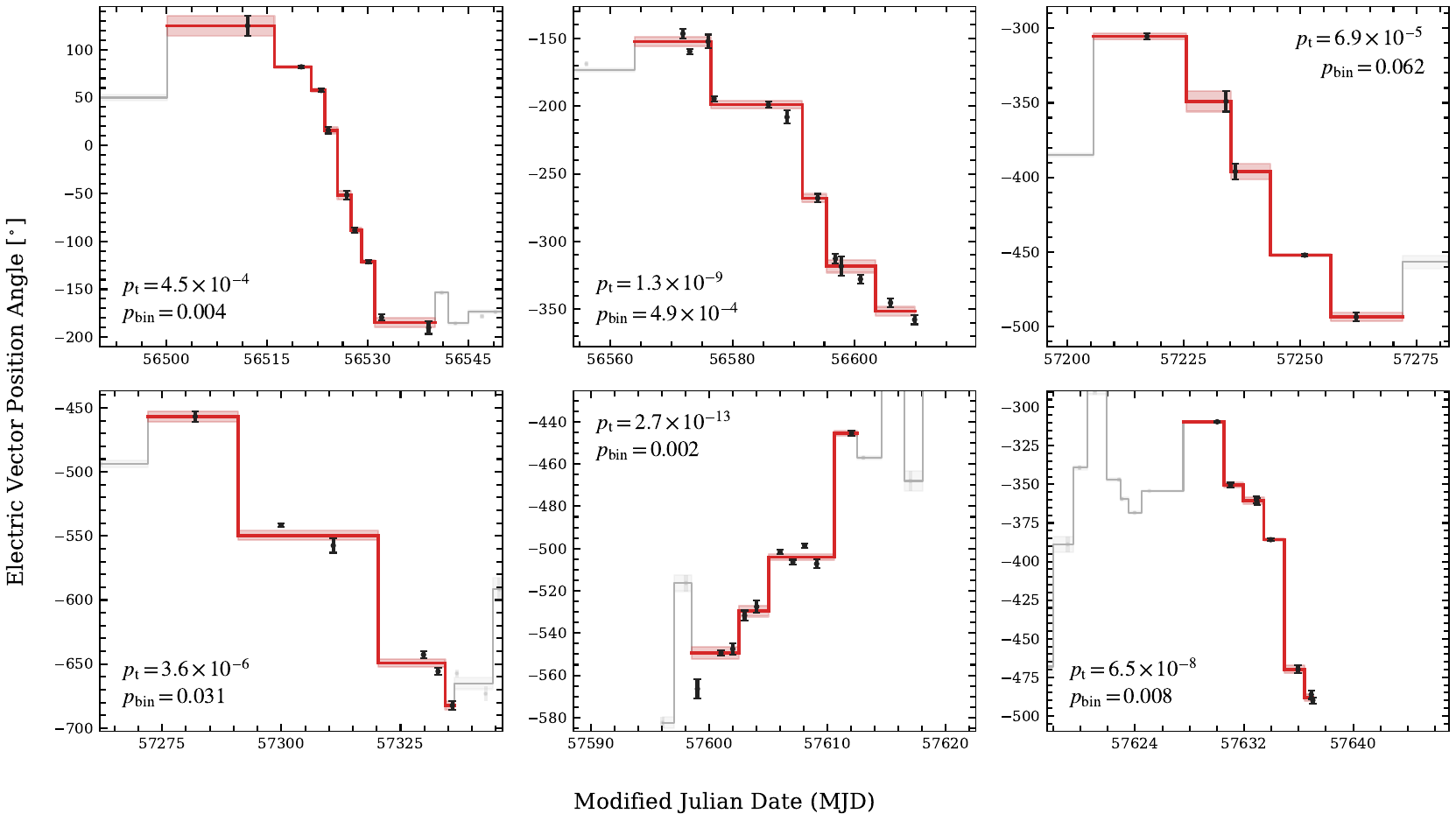}

    \caption{Detected EVPA rotation events for CTA~102 (RBPLJ2232+1143). Each panel corresponds to an individual rotation interval and shows the EVPA evolution as a function of MJD together with the Bayesian Blocks trend (red line) and the observed data points with uncertainties.The middle panel in the bottom row illustrates a case that is included by our method because the extrema pair satisfies the amplitude and statistical thresholds, even though the rotation was interrupted at the second point of the third Bayesian Block. Under previous smoothness-based criteria, this interruption would result in two distinct rotations, neither of which satisfies the $90^\circ$ criterion.}
    \label{fig:rotations}
\end{figure*}

The identification of EVPA rotation events was performed with a dedicated pipeline that integrates Bayesian Blocks segmentation, local-extrema analysis, and statistical validation.
In contrast to the previous EVPA identification method used by RoboPol, our method does not impose smoothness constraints on the EVPA curve; candidate rotations are identified
solely on the basis of statistically significant extrema pairs, regardless of the derivative behavior. {As a result, events which would not have qualified as rotations under earlier criteria are included here. An example is shown in the middle panel of the bottom row of
Fig.~\ref{fig:rotations}: the rotation was interrupted at the second point of the third Bayesian Block, meaning the previous method would have split it into two distinct rotations that both fail to satisfy the $90^\circ$ criterion.}

\begin{figure*}[htp!]
    \centering
    \includegraphics[width=0.9\linewidth]{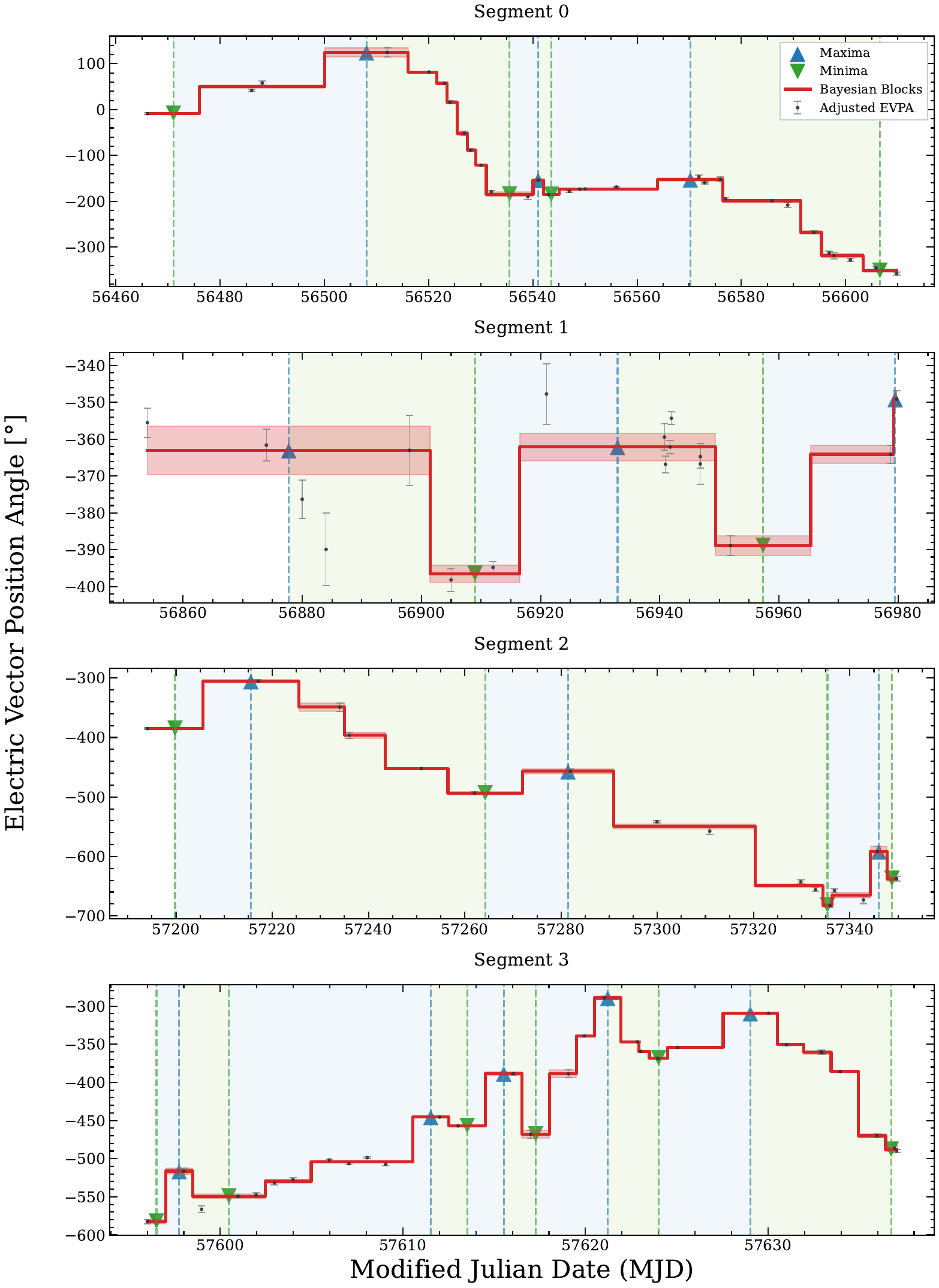}
    \caption{Bayesian Blocks segmentation of the EVPA time series for CTA~102 (RBPLJ2232+1143).
    The panels show the accepted {segments}, defined by the criteria of more than four data points and a maximum spacing of 30 days between consecutive observations. Within each {segment}, the Bayesian Blocks step representation is overplotted, and the global minima and maxima are marked with triangles.}
    \label{fig:segments}
\end{figure*}

As a first step, the time series of each source was divided into contiguous segments, which were retained only if they contained more than four data points and no pair of consecutive points was separated by more than 30 days. This segmentation is illustrated in Fig.~\ref{fig:segments}, where each accepted {segment} is shown together with its Bayesian Blocks step representation and the global minima and maxima are marked with triangles.

Within each accepted segment, Bayesian Blocks were applied to obtain block means and
uncertainties, from which local maxima and minima were extracted. Local extrema were
identified using the \texttt{argrelextrema} function from the \texttt{scipy.signal} Python library.

The procedure also explicitly checks the first and last bins to ensure that boundary points are included if they represent extrema. The resulting sets of maxima and minima are stored together with their corresponding MJDs.

Adjacent extrema pairs of type max$\rightarrow$min or min$\rightarrow$max were treated as candidate rotations, with their temporal boundaries defined by the corresponding block edges. Only candidate rotations with an EVPA amplitude exceeding $90^{\circ}$ were retained, ensuring that only large variations are considered.

{Candidate rotation intervals} were retained only if the sub-segmentation comprised three or more {Bayesian Blocks}, ensuring that detected rotations exhibit sufficient internal structure. In addition, each candidate interval was required to contain at least four data points, so that subsequent statistical tests could be meaningfully applied and spurious fluctuations avoided.

Each candidate interval was then subjected to statistical tests: a one-sample $t$-test to assess deviation from a reference mean, and a binomial test to quantify asymmetry in the EVPA distribution.
The rotation amplitude was computed as the difference between maximum and minimum EVPA values, with uncertainties propagated from the measurement errors. The duration ($\Delta$MJD) and its uncertainty were derived from the block boundaries, enabling computation of the rotation rate and its uncertainty. Only events exceeding the minimum amplitude threshold and passing all statistical criteria were accepted as rotations. Finally, the pipeline summarizes results across sources {(giving the number of detected rotation events, per-source distributions, and descriptive statistics for amplitudes and duration)}, and produces plots for each source.

\section{Results}
\subsection{EVPA rotation catalog and summary statistics}

The full catalog of EVPA rotation events detected by our pipeline is provided in Appendix~A (Table~\ref{Appendix1}). The catalog comprises $N=48$ detected rotations from 25 unique RoboPol sources; several objects show multiple events (e.g., RBPLJ2232+1143: 6 events;RBPLJ1751+0939, RBPLJ1800+7828, RBPLJ2253+1608: 4 events each). The rotation amplitude distribution has a mean of  $\langle\Delta\theta\rangle = 159.7^{\circ}$ (median $=153.8^{\circ}$, and $\sigma = 66.5^{\circ}$); the largest recorded amplitude is $359.7^{\circ}$ (RBPLJ1751+0939) and the smallest is $90.8^{\circ}$ (RBPLJ1635+3808). Event durations span $7.0$--$111.3$ days (mean
$=44.4$ days, median $=42.4$ days, $\sigma = 23.8$ days). The mean rotation rate is
$5.0^{\circ}$/day (median $=3.9^{\circ}$/day).

\subsection{Comparison with previous RoboPol results}

We cross-matched EVPA rotations reported by \citet{blinov2015,blinov2016b,blinov2016a,Blinov2018} with those identified using our Bayesian Blocks approach. For each overlapping event, we quantified the discrepancy in units of the Bayesian uncertainty by computing the difference in sigma

\begin{equation}
      \Delta = \frac{|x - y|}{\sigma_x}
      \label{eq:delta}
\end{equation}

\noindent where $x$ is the Bayesian Blocks measurement, $y$ the catalog value, and $\sigma_x$ the Bayesian uncertainty of each parameter. In this definition, $\Delta$ (Eq.~\ref{eq:delta}) expresses how many standard deviations separate the two measurements. Differences below $3\sigma$ were considered consistent, and importantly, more than half of the overlapping rotation events fall within this threshold, confirming that the majority of Bayesian Blocks detections are statistically compatible
with the RoboPol catalog values.

\begin{figure*}[!ht]
    \centering
    \includegraphics[width=\textwidth]{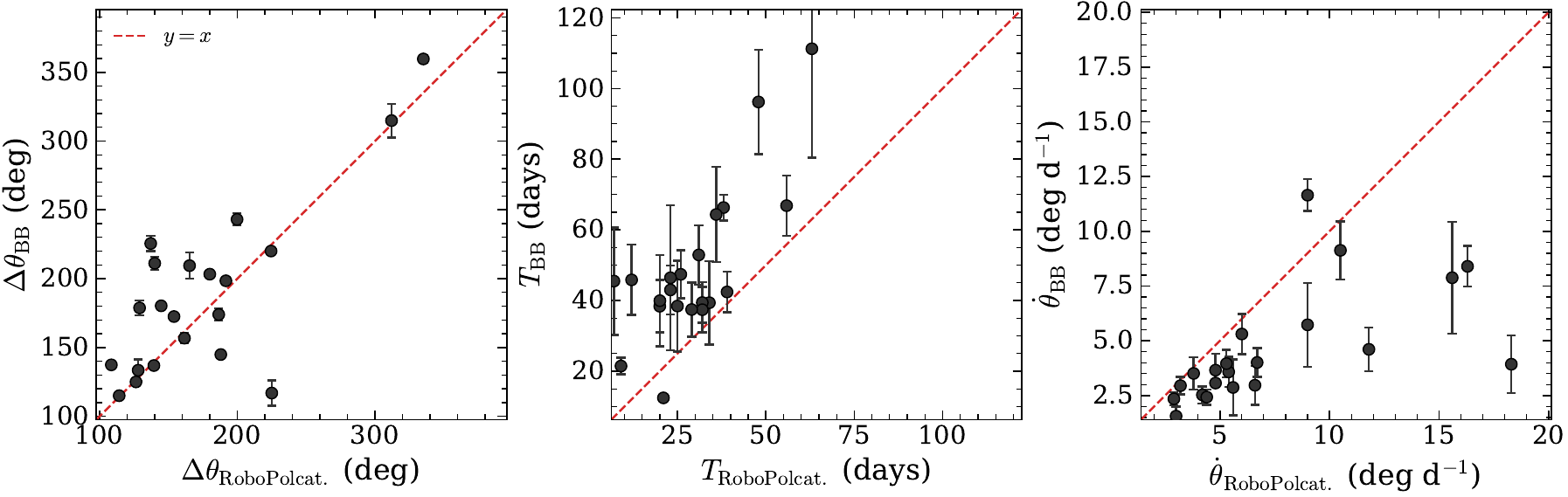}
   \caption{Scatter comparison of EVPA rotation parameters between the RoboPol catalogs of
    \citet{blinov2015,blinov2016b,blinov2016a,Blinov2018} (x-axis, catalogue values)
    and the Bayesian Blocks results (y-axis).
    Panels show amplitude ($\Delta\theta_{\max}$), rotation period ($T_{\rm rot}$), and angular
    velocity ({$\dot{\theta}$}). Each point represents an overlapping rotation event. The red dashed line marks the identity relation $y = x$. A systematic offset is evident in the rotation period panel, with durations tending to be larger in the Bayesian Blocks catalog. The slower angular velocities reflect this difference in duration, while amplitudes show no strong systematic offset.}
    \label{fig:scatter_comparison}
\end{figure*}

Fig.~\ref{fig:scatter_comparison} presents scatter plots comparing amplitude, rotation period, and angular velocity between the two methods. Each point corresponds to an overlapping rotation event. The red dashed line marks the identity relation $y = x$.

A systematic offset is evident in the rotation period panel: durations tend to be
longer in the Bayesian Blocks catalog than in the RoboPol catalog. The slower angular velocities reflect this difference in duration. Amplitudes, by contrast, do not show a strong systematic offset between the two methods.

\begin{figure*}[!ht]
    \centering
    \includegraphics[width=\textwidth]{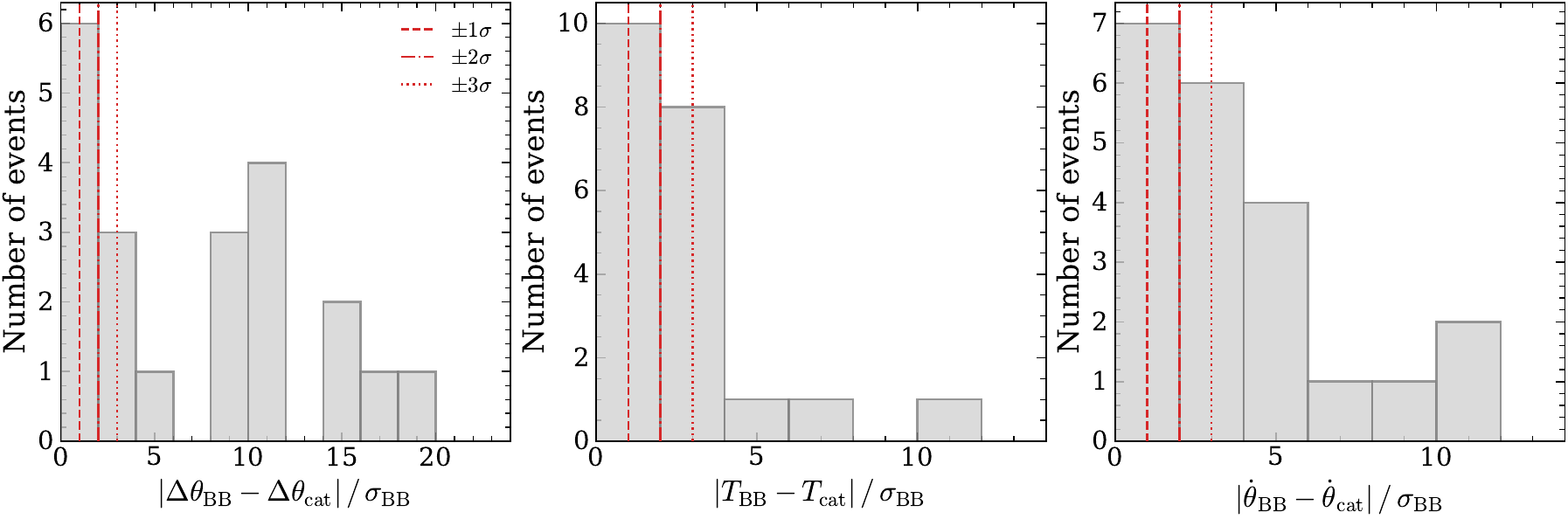}
    \caption{Distributions of differences in sigma for overlapping EVPA rotation events. Panels show (a) amplitude, (b) rotation period, and (c) angular velocity. Histograms display normalized events of $\Delta = |x - y|/\sigma_x$ for each parameter. Vertical red lines indicate  $\pm1\sigma$ (dashed ), $\pm2\sigma$ (dashed dotted), and $\pm3\sigma$ ( dotted) thresholds.
    Amplitude differences (panel a) are broadly distributed with a main peak near
    zero and a secondary peak at large $\Delta$ values; period differences (panel b) are skewed positive; and angular velocity differences (panel c) differences are primarily clustered within 4$\sigma$ with minor outliers at higher $\Delta$.}
    \label{fig:residual_histograms}
\end{figure*}

Fig.~\ref{fig:residual_histograms} shows the corresponding distributions of differences in sigma. 
Amplitude differences (panel a) are broadly distributed with a main peak near zero, 
indicating overall consistency between the two methods, though a secondary peak at 
large $\Delta$ values ($>10\sigma$) reflects a minority of strongly discrepant events.

Period differences (panel b) and angular velocity differences (panel c) are both positively skewed,  with the majority of events clustered within $4\sigma$ and a few outliers at higher $\Delta$.  These distributions confirm that the discrepancies follow a non-Gaussian, systematic pattern  rather than purely random noise.

Overall, Bayesian Blocks rotations are on average $\sim10\%$ larger in amplitude, $\sim2\times$ longer in duration, and $\sim2/3$ slower in angular velocity compared to the RoboPol catalog. While correlation coefficients remain moderate to strong, the sigma differences highlight methodological biases introduced by block segmentation. Specifically, the block-based approach extends rotation intervals until the next significant EVPA variation, which can inflate durations and suppress angular velocities. Despite these differences, the detection of rotations remains robust, and the discrepancies reflect definitional choices rather than statistical inconsistencies.

\subsection{Additional unreported rotations}
Our Bayesian Blocks analysis revealed a total of 27 EVPA rotations that were not included in the Blinov et al. catalog. These comprise two categories:

\begin{itemize}
\item{\textbf{Earlier seasons (2013--2016).}
Within the monitoring intervals already covered by
\cite{blinov2015,blinov2016b,blinov2016a,Blinov2018}, we detect 16 additional rotations that were not reported previously. These events occur in the same seasons as
the published rotations but were missed by the original procedure. Their parameters (amplitude,
duration, angular velocity) are listed in Table~\ref{tab:unreported_rotations}, together with
uncertainties derived from the Bayesian Blocks method.}

\item{\textbf{Final season (2016--2017).}
In addition, our analysis extends beyond the
\cite{blinov2015,blinov2016b,blinov2016a,Blinov2018} detections to the final observin season. Here we detect 11 new rotations that expand the temporal coverage of the sample. Together, these 27 unreported events demonstrate that the Bayesian Blocks method provides a more complete inventory of EVPA rotations
across the full monitoring period, both by recovering overlooked events in earlier seasons and by
extending the catalog into new temporal territory.}
\end{itemize}

\subsection{Correlation with \texorpdfstring{$\gamma$}{gamma}--ray activity}

\begin{table*}[!ht]
\caption{Summary of 27 unreported EVPA rotations detected with Bayesian Blocks.
The sample includes 16 events from seasons already covered by
\citet{blinov2015,blinov2016b,blinov2016a,Blinov2018} and 11 new events from the
final 2016--2017 season. Parameters are reported with Bayesian uncertainties, shown as value $\pm$
error.}
\label{tab:unreported_rotations}
\centering
\begin{tabular}{l l l c c c c}
\toprule
Season & Blazar ID & Simbad Name & Spectral Type & Amp [$^\circ$] & Period [days] & Ang. vel. [$^\circ$/day] \\
\midrule
\multicolumn{7}{c}{Unreported rotations in 2013--2016 (16 events)} \\
\midrule
2015 & RBPLJ1505+0326 & PMN J1505+0326     & ISP & $126.7 \pm 7.9$ & $87.9 \pm 14.0$ & $1.4 \pm 0.3$ \\
2015 & RBPLJ1512-0905 & PKS 1510-089       & LSP & $93.9  \pm 3.4$ & $64.4 \pm 11.3$ & $1.5 \pm 0.3$ \\
2015 & RBPLJ1542+6129 & QSO B1542+614      & LSP & $109.3 \pm 4.7$ & $61.9 \pm 30.5$ & $1.8 \pm 0.9$ \\
2015 & RBPLJ1555+1111 & PG 1553+113        & HSP & $142.1 \pm 4.7$ & $38.4 \pm 7.3$  & $3.7 \pm 0.8$ \\
2015 & RBPLJ1558+5625 & TXS 1557+565       & ISP & $115.0 \pm 5.7$ & $94.4 \pm 19.0$ & $1.2 \pm 0.3$ \\
2015 & RBPLJ1604+5714 & 7C 1603+5722       & ISP & $195.9 \pm 6.5$ & $48.0 \pm 11.9$ & $4.1 \pm 1.1$ \\
2015 & RBPLJ1725+1152 & H 1722+119         & HSP & $97.3  \pm 6.4$ & $21.0 \pm 5.3$  & $4.6 \pm 1.4$ \\
2015 & RBPLJ1751+0939 & OT 081 & LSP & $359.7 \pm 2.8$ & $39.4 \pm 5.8$ & $9.1 \pm 1.4$ \\
2016 & RBPLJ1751+0939 & OT 081             & LSP & $134.8 \pm 3.1$ & $44.4 \pm 11.0$ & $3.0 \pm 0.8$ \\
2015 & RBPLJ1800+7828 & S5 1803+78         & LSP & $111.1 \pm 2.4$ & $81.9 \pm 32.2$ & $1.4 \pm 0.6$ \\
2016 & RBPLJ1838+4802 & ATO J279.7048+48.0428 & HSP & $159.8 \pm 5.7$ & $97.4 \pm 20.1$ & $1.6 \pm 0.4$ \\
2015 & RBPLJ2022+7611 & S5 2023+76         & ISP & $136.8 \pm 2.3$ & $58.9 \pm 5.0$  & $2.3 \pm 0.2$ \\
2015 & RBPLJ2149+0322 & MITG J2149+0323    & ISP & $94.0  \pm 3.3$ & $57.3 \pm 8.5$  & $1.6 \pm 0.3$ \\
2016 & RBPLJ2232+1143 & CTA 102            & LSP & $188.3 \pm 3.6$ & $66.4 \pm 17.9$ & $2.8 \pm 0.8$ \\
2016 & RBPLJ1754+3212 & ATO J268.5491+32.2064 & ISP & $165.9 \pm 3.8$ & $67.9 \pm 11.9$ & $2.4 \pm 0.5$ \\
2016 & RBPLJ1754+3212 & ATO J268.5491+32.2064 & ISP & $99.5  \pm 4.0$ & $50.4 \pm 7.6$  & $2.0 \pm 0.3$ \\
\midrule
\multicolumn{7}{c}{Unreported rotations in 2016--2017 (11 events)} \\
\midrule
2016 & RBPLJ1512-0905 & PKS 1510-089       & LSP & $112.4 \pm 4.8$  & $14.5 \pm 2.4$  & $7.8 \pm 1.7$  \\
2016 & RBPLJ1635+3808 & 4C +38.41          & LSP & $153.5 \pm 5.0$  & $24.5 \pm 8.2$  & $6.3 \pm 2.2$  \\
2016 & RBPLJ1635+3808 & 4C +38.41          & LSP & $90.8  \pm 6.7$  & $7.5  \pm 2.3$  & $12.2 \pm 4.3$ \\
2016 & RBPLJ1754+3212 & ATO J268.5491+32.2064 & ISP & $119.8 \pm 10.8$ & $22.0 \pm 2.9$  & $5.5 \pm 1.0$  \\
2016 & RBPLJ1800+7828 & S5 1803+78         & LSP & $194.8 \pm 3.9$  & $18.0 \pm 7.3$  & $10.8 \pm 4.6$ \\
2016 & RBPLJ2022+7611 & S5 2023+76         & ISP & $164.7 \pm 1.8$  & $13.6 \pm 2.1$  & $12.1 \pm 2.0$ \\
2016 & RBPLJ2202+4216 & BL Lacertae        & LSP & $104.3 \pm 3.2$  & $14.4 \pm 4.4$  & $7.2 \pm 2.3$  \\
2016 & RBPLJ2202+4216 & BL Lacertae        & LSP & $130.7 \pm 4.4$  & $18.1 \pm 1.5$  & $7.2 \pm 0.8$  \\
2016 & RBPLJ2232+1143 & CTA 102            & LSP & $121.0 \pm 4.5$  & $14.0 \pm 3.1$  & $8.6 \pm 2.1$  \\
2016 & RBPLJ2232+1143 & CTA 102            & LSP & $180.5 \pm 2.1$  & $9.5  \pm 2.2$  & $19.0 \pm 4.5$ \\
2016 & RBPLJ2253+1608 & 3C 454.3           & LSP & $167.9 \pm 4.2$  & $7.0  \pm 2.1$  & $23.8 \pm 7.6$ \\

\bottomrule
\end{tabular}
\end{table*}

We investigated the correlation between optical EVPA rotations and $\gamma$--ray activity using all 48 detected rotation events (Table~\ref{tab:enhanced_rotations}).

For this purpose, we retrieved $\gamma$--ray light curves from the \textit{Fermi}{--LAT} repository \citep{Abdollahi_2023}. The light curves were constructed using a standard binning of 7~days, which provides sufficient temporal resolution while maintaining adequate photon statistics. For each source, we extracted the energy flux in the 0.1--300~GeV band, reported in units of MeV\,cm$^{-2}$\,s$^{-1}$, as a function of {MJD}.

We then examined whether enhanced $\gamma$--ray activity coincides with the time intervals during which EVPA rotations are observed. To quantify this, we calculate the peak-to-mean $\gamma$--ray energy flux from the 7-day binned \textit{Fermi}--LAT light curve by dividing the maximum flux within the rotation interval by the mean flux over the same period interval. We then compare that to the amplitude and duration of the rotation using the Spearman rank correlation coefficient. This test allows us to assess whether stronger or longer rotations are statistically associated with increased
$\gamma$--ray activity.

\begin{figure*}[!ht]
  \centering
  \includegraphics[width=\textwidth]{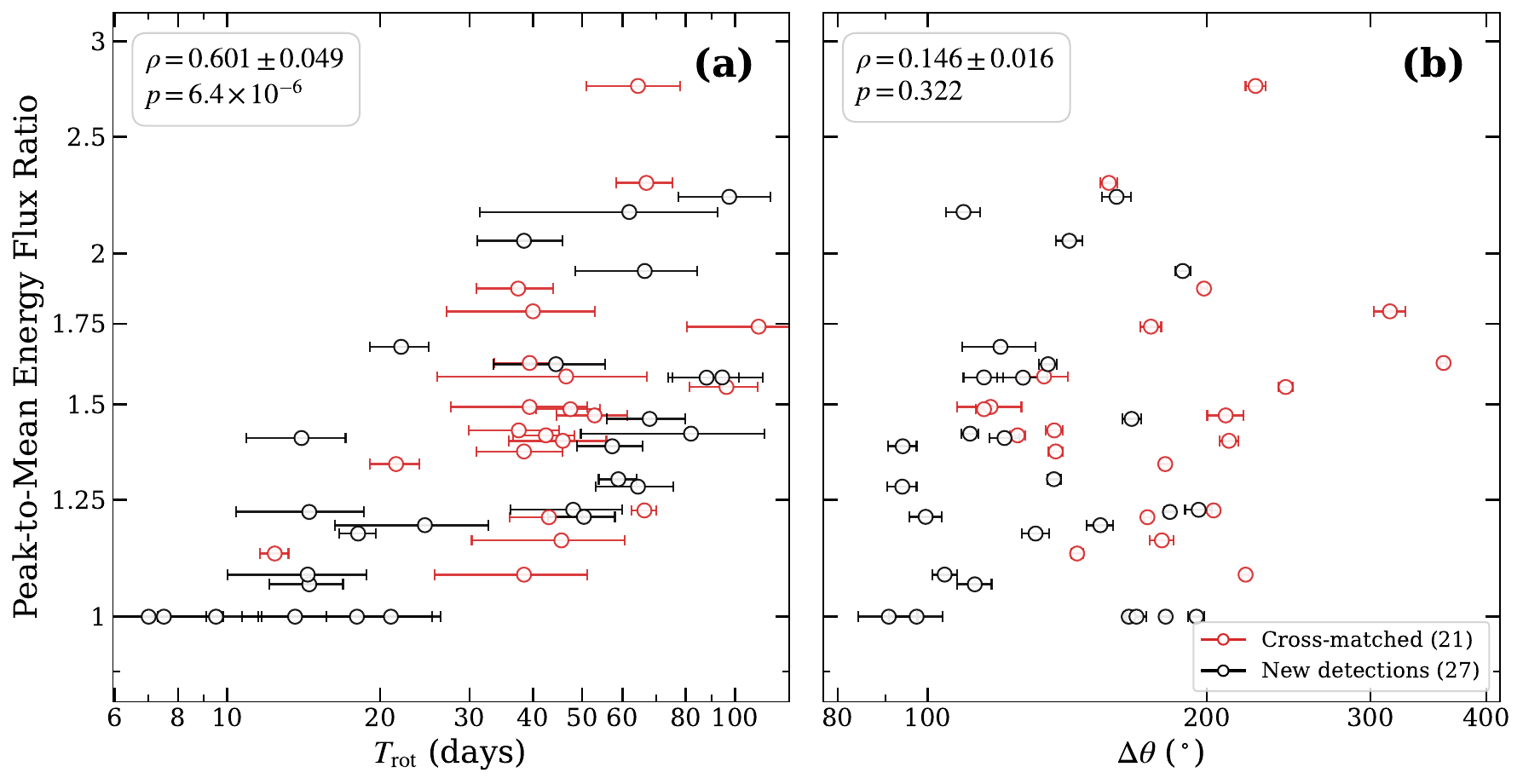}
  \caption{Spearman correlations between the \textit{Fermi}--LAT peak-to-mean $\gamma$-ray energy flux and rotation parameters for all 48 detected rotation events, split into events overlapping with the \citet{blinov2015,blinov2016b,blinov2016a,Blinov2018} catalog (red circles, $n=21$) and the 27 newly reported events (black circles, $n=27$).
  (a) Rotation period $T_{\mathrm{rot}}$ versus peak-to-mean ratio.
  (b) Maximum EVPA amplitude $\Delta\theta_{\mathrm{max}}$ versus peak-to-mean ratio.
  Each panel reports the Spearman rank coefficient $\rho$ with its Monte Carlo
  uncertainty, the associated $p$-value computed over all 48 events, and the
  sample size $n$ as annotated. Axes are in log--log scale; horizontal error bars indicate uncertainties on the x-axis parameters.}
  \label{fig:gamma_spearman_period_amplitude}
\end{figure*}

The sample comprises all 48 detected rotation events (Table~\ref{tab:enhanced_rotations}), of which 21 overlap with the
\citet{blinov2015,blinov2016b,blinov2016a,Blinov2018} catalog and 27 are newly reported here.
We use the Spearman rank coefficients because the test is robust to nonlinearity and outliers;uncertainties on the rotation parameters are shown as horizontal error bars in
Fig.~\ref{fig:gamma_spearman_period_amplitude} but were not propagated into the correlation test.
To assess the robustness of the correlation against measurement uncertainties, we
performed a Monte Carlo analysis with $N_{\rm iter} = 50\,000$ realisations. In each realisation, every rotation parameter ($T_{\rm rot}$ or $\Delta\theta_{\rm max}$) was independently perturbed by drawing from a Gaussian distribution centred on its measured value with standard deviation equal to its measurement uncertainty. The Spearman rank correlation coefficient $\rho$ and its associated $p$-value were then recomputed for the full perturbed sample of 48 events. This yielded a distribution of $\rho$ values from which the median and standard deviation were derived; the median $\rho$ and $\sigma_\rho$ are reported in Fig.~\ref{fig:gamma_spearman_period_amplitude}. We also
record the fraction of realisations in which the $p$-value falls below significance thresholds of 0.05, 0.01, and 0.001, providing a direct measure of how robust the correlation is against parameter uncertainties. All 50\,000 realisations produced valid Spearman estimates (no failed iterations).
Panel~(a) of Fig.~\ref{fig:gamma_spearman_period_amplitude} shows a significant positive correlation between rotation period and peak-to-mean $\gamma$-ray flux
($\rho = 0.601 \pm 0.049$, $p_{\rm med} = 6.4\times10^{-6}$, $n=48$).
The correlation is highly robust: 100\% of MC realisations yield $p < 0.05$,
confirming that longer rotations tend to coincide with stronger $\gamma$-ray enhancements regardless of measurement uncertainties.
In contrast, panel~(b) shows no significant correlation between maximum EVPA amplitude
$\Delta\theta_{\mathrm{max}}$ and peak-to-mean flux
($\rho = 0.146 \pm 0.016$, $p_{\rm med} = 0.322$, $n=48$).
None of the 50\,000 MC realisations yields $p < 0.05$, confirming that rotation
amplitude alone is not predictive of contemporaneous $\gamma$-ray brightness.
\section{Conclusions}

We have developed and applied an automated pipeline for the detection of EVPA rotations in blazars, combining error-weighted angle adjustment, Bayesian Blocks segmentation, and statistical validation.
Applied to RoboPol monitoring data, the method identified $N=48$ rotations across 25 sources, including multiple events in RBPLJ2232+1143 (6), RBPLJ1751+0939, RBPLJ1800+7828, and RBPLJ2253+1608 (4 each). The cataloged rotations span amplitudes from $90.8^{\circ}$ to $359.7^{\circ}$, with a mean of $159.7^{\circ} \pm 66.5^{\circ}$, durations between $7.0$ and $111.3$ days (mean $44.4 \pm 23.8$ days), and rotation rates averaging $5.01^{\circ}$/day.

Comparison with previous catalogs \cite{blinov2015,blinov2016b,blinov2016a,Blinov2018} reveals systematic differences: Bayesian Blocks rotations are on average $\sim 10\%$ larger in amplitude, $\sim 2\times$ longer in duration, and $\sim 2/3$ slower in angular velocity. These offsets reflect methodological biases, with adaptive binning merging adjacent swings into longer events, in contrast to the original RoboPol segmentation methodology.

In addition, our analysis uncovered 27 previously unreported rotations, including 16 events from seasons already covered by Blinov et al. and 11 new events from the final 2016--2017 season. These rotations span amplitudes of $90.8^{\circ}$--$359.7^{\circ}$, durations of 7--97 days, and angular velocities of $1.2$--$23.8^{\circ}$/day, further enriching the statistical sample of EVPA variability.
A Monte Carlo correlation analysis ($N_{\rm iter}=50\,000$) with contemporaneous
Fermi--LAT $\gamma$-ray light curves, carried out over all 48 detected rotation events, shows that longer rotations tend to coincide with stronger $\gamma$-ray enhancements
($\rho = 0.601 \pm 0.049$, $p_{\rm med} = 6.4\times10^{-6}$; significant in 100\% of MC
realisations at $p<0.05$), while rotation amplitude alone is not predictive of $\gamma$-ray brightness ($\rho = 0.146 \pm 0.016$, $p_{\rm med} = 0.322$; significant in 0\% of realisations at $p<0.05$).

Beyond the statistical characterization, the detected rotations carry important physical
implications. The tendency for longer EVPA rotations to coincide with enhanced $\gamma$-ray activity suggests that high-energy emission is linked to sustained reconfigurations of the jet magnetic field, indicating the presence of long-lived polarized components in the jet. On the other hand, the lack of correlation between $\gamma$-ray activity and rotation amplitude implies the presence of a stochastic component as well. These results highlight the role of polarization monitoring as a probe of jet dynamics and particle acceleration in blazars.

Our pipeline minimizes subjective biases, ensures reproducible identification of EVPA rotations, and provides a statistically rigorous framework for cataloging polarization variability. Future applications to larger samples and multiwavelength datasets \citep[e.g., X-ray polarization angle rotations,][]{DiGesu2023,Maksym2025}, including contemporaneous $\gamma$-ray activity, will further constrain the physical mechanisms driving rotations.

\section*{Data and software availability}

The data supporting the findings of this article are openly available in the Harvard Dataverse \citep{Blinov2021}.

The analysis pipeline developed for this work is openly available at
\href{https://github.com/glykanastasia/blazar-evpa-rotation-detector}{https://github.com/glykanastasia/blazar-evpa-rotation-detector}.
It provides all scripts used for EVPA rotation detection, Bayesian Blocks segmentation, and figure generation, ensuring full reproducibility of the results presented in this paper.

\begin{acknowledgements}
      We thank the anonymous referee for their careful reading of the manuscript and their constructive comments, which helped improve the quality of this work. A.G. and I.L were funded by the European Union ERC-2022-STG - BOOTES - 101076343. Views and opinions expressed are however those of the author(s) only and do not necessarily reflect those of the European Union or the European Research Council Executive Agency. Neither the European Union nor the granting authority can be held responsible for them. D.B. acknowledges support from the European Research Council (ERC) under the Horizon ERC Grants 2021 programme under grant agreement No. 101040021
\end{acknowledgements}

\bibliographystyle{bibtex/aa}
\bibliography{references}
\onecolumn

\appendix
\section{Catalog of detected EVPA rotations}
\label{Appendix1}
\begingroup
\tiny
\setlength{\tabcolsep}{3pt}
\renewcommand{\arraystretch}{0.98}
\setlength{\LTcapwidth}{\textwidth}

\begin{center}
  \makebox[\textwidth][c]{
    \rotatebox{90}{
      \begin{minipage}{\textheight}
        \begin{longtable}{@{}llllrrrrccr@{}}
        \caption{EVPA rotation events detected in the RoboPol 2013--2017 dataset. The table lists
        Simbad identifiers, \textit{Fermi}-LAT counterparts, and spectral classifications. All events have
        amplitudes $\Delta\theta \geq 90^\circ$ and satisfy the adopted statistical significance
        criteria ($p<0.05$ for t-test, $p<0.065$ for binomial test).}
        \label{tab:enhanced_rotations} \\
        \toprule
        \textbf{Source} & \textbf{Simbad Name} & \textbf{\textit{Fermi}-LAT Name} & \textbf{Spectral Type} & \textbf{Amplitude (\textdegree)} & \textbf{Start (MJD)} & \textbf{End (MJD)} & \textbf{$\Delta$MJD (days)} & \textbf{p-value (t-test)} & \textbf{p-value (binomial)} & \textbf{Rate (\textdegree/day)} \\
        \midrule
        \endfirsthead

        \multicolumn{11}{c}%
        {{\bfseries \tablename\ \thetable{} -- continued from previous page}} \\
        \toprule
        \textbf{Source} & \textbf{Simbad Name} & \textbf{\textit{Fermi}-LAT Name} & \textbf{Spectral Type} & \textbf{Amplitude (\textdegree)} & \textbf{Start (MJD)} & \textbf{End (MJD)} & \textbf{$\Delta$MJD (days)} & \textbf{p-value (t-test)} & \textbf{p-value (binomial)} & \textbf{Rate (\textdegree/day)} \\
        \midrule
        \endhead

        \midrule \multicolumn{11}{@{}r@{}}{{Continued on next page}} \\
        \endfoot

        \bottomrule
        \endlastfoot
        RBPLJ0045+2127  & GB6 J0045+2127 & 4FGL J0045.3+2128 & HSP & $243.1 \pm 4.4$  & 57253.6 & 57349.8 & $96.2 \pm 14.8$  & 1.4$\times10^{-2}$ & 6.3$\times10^{-3}$ & $2.5 \pm 0.4$ \\
        RBPLJ0136+4751  & OC 457 & 4FGL J0137.0+4751 & LSP & $116.9 \pm 9.3$  & 56561.1 & 56600.4 & $39.4 \pm 11.8$  & 2.9$\times10^{-3}$ & 3.1$\times10^{-2}$ & $3.0 \pm 1.0$ \\
        RBPLJ0136+4751  & OC 457 & 4FGL J0137.0+4751 & LSP & $115.0 \pm 2.4$  & 57264.0 & 57311.5 & $47.4 \pm 6.8$   & 1.7$\times10^{-3}$ & 3.1$\times10^{-2}$ & $2.4 \pm 0.4$ \\
        RBPLJ0136+4751  & OC 457 & 4FGL J0137.0+4751 & LSP & $137.4 \pm 2.5$  & 57311.5 & 57349.8 & $38.4 \pm 7.4$   & 1.6$\times10^{-8}$ & 2.0$\times10^{-3}$ & $3.6 \pm 0.7$ \\
        RBPLJ0259+0747  & PKS 0256+075 & 4FGL J0259.4+0746 & LSP & $203.3 \pm 2.1$  & 56543.6 & 56609.9 & $66.3 \pm 3.7$   & 2.8$\times10^{-4}$ & 7.8$\times10^{-3}$ & $3.1 \pm 0.2$ \\
        RBPLJ0854+2006  & OJ 287 & 4FGL J0854.8+2006 & LSP & $172.5 \pm 1.6$  & 56556.1 & 56599.1 & $43.0 \pm 7.0$   & 4.5$\times10^{-9}$ & 4.9$\times10^{-4}$ & $4.0 \pm 0.7$ \\
        RBPLJ1037+5711  & 87GB 103431.3+572750 & 4FGL J1037.7+5711 & ISP & $209.5 \pm 9.4$  & 56776.9 & 56829.8 & $52.9 \pm 8.4$   & 6.4$\times10^{-5}$ & 1.6$\times10^{-2}$ & $4.0 \pm 0.8$ \\
        RBPLJ1048+7143  & S5 1044+71 & 4FGL J1048.4+7143 & LSP & $144.9 \pm 2.2$  & 56594.6 & 56607.1 & $12.4 \pm 0.8$   & 1.5$\times10^{-9}$ & 9.8$\times10^{-4}$ & $11.7 \pm 0.9$ \\
        RBPLJ1505+0326  & PMN J1505+0326 & 4FGL J1505.0+0326 & ISP & $126.7 \pm 7.9$  & 56773.9 & 56861.8 & $87.9 \pm 14.0$  & 1.1$\times10^{-4}$ & 6.3$\times10^{-2}$ & $1.4 \pm 0.3$ \\
        RBPLJ1512-0905  & PKS 1510-089 & 4FGL J1512.8-0906 & LSP & $93.9 \pm 3.4$   & 56802.4 & 56866.8 & $64.4 \pm 11.3$  & 1.8$\times10^{-2}$ & 6.3$\times10^{-2}$ & $1.5 \pm 0.3$ \\
        RBPLJ1512-0905  & PKS 1510-089 & 4FGL J1512.8-0906 & LSP & $112.4 \pm 4.8$  & 57604.8 & 57619.3 & $14.5 \pm 2.4$   & 2.4$\times10^{-5}$ & 1.6$\times10^{-2}$ & $7.8 \pm 1.7$ \\
        RBPLJ1542+6129  & QSO B1542+614 & 4FGL J1543.0+6130 & LSP & $109.3 \pm 4.7$  & 56797.0 & 56858.8 & $61.9 \pm 30.5$  & 1.4$\times10^{-6}$ & 7.8$\times10^{-3}$ & $1.8 \pm 0.9$ \\
        RBPLJ1555+1111  & PG 1553+113 & 4FGL J1555.7+1111 & HSP & $133.5 \pm 8.0$  & 56499.3 & 56545.8 & $46.5 \pm 20.6$  & 1.1$\times10^{-8}$ & 9.8$\times10^{-4}$ & $2.9 \pm 1.4$ \\
        RBPLJ1555+1111  & PG 1553+113 & 4FGL J1555.7+1111 & HSP & $142.1 \pm 4.7$  & 56776.5 & 56814.9 & $38.4 \pm 7.3$   & 4.8$\times10^{-7}$ & 4.9$\times10^{-4}$ & $3.7 \pm 0.8$ \\
        RBPLJ1558+5625  & TXS 1557+565 & 4FGL J1558.8+5625 & ISP & $115.0 \pm 5.7$  & 57168.4 & 57262.8 & $94.4 \pm 19.0$  & 1.1$\times10^{-10}$ & 3.9$\times10^{-3}$ & $1.2 \pm 0.3$ \\
        RBPLJ1604+5714  & 7C 1603+5722 & 4FGL J1604.6+5714 & ISP & $195.9 \pm 6.5$  & 57187.9 & 57235.8 & $48.0 \pm 11.9$  & 1.7$\times10^{-2}$ & 6.3$\times10^{-2}$ & $4.1 \pm 1.1$ \\
        RBPLJ1635+3808  & 4C +38.41 & 4FGL J1635.2+3808 & LSP & $153.5 \pm 5.0$  & 57587.9 & 57612.4 & $24.5 \pm 8.2$   & 7.5$\times10^{-6}$ & 3.1$\times10^{-5}$ & $6.3 \pm 2.2$ \\
        RBPLJ1635+3808  & 4C +38.41 & 4FGL J1635.2+3808 & LSP & $90.8 \pm 6.7$   & 57625.9 & 57633.3 & $7.5 \pm 2.3$    & 4.4$\times10^{-4}$ & 6.3$\times10^{-2}$ & $12.2 \pm 4.3$ \\
        RBPLJ1725+1152  & H 1722+119 & 4FGL J1725.0+1152 & HSP & $97.3 \pm 6.4$   & 56523.8 & 56544.8 & $21.0 \pm 5.3$   & 3.8$\times10^{-4}$ & 3.1$\times10^{-2}$ & $4.6 \pm 1.4$ \\
        RBPLJ1748+7005  & S5 1749+70 & 4FGL J1748.6+7005 & LSP & $125.0 \pm 2.4$  & 56854.9 & 56897.3 & $42.4 \pm 5.8$   & 1.2$\times10^{-12}$ & 1.2$\times10^{-4}$ & $3.0 \pm 0.4$ \\
        RBPLJ1751+0939  & OT 081 & 4FGL J1751.5+0938 & LSP & $182.4 \pm 2.4$  & 56524.3 & 56538.8 & $14.5 \pm 4.1$   & 5.6$\times10^{-3}$ & 6.3$\times10^{-2}$ & $12.6 \pm 3.7$ \\
        RBPLJ1751+0939  & OT 081 & 4FGL J1751.5+0938 & LSP & $359.7 \pm 2.8$  & 56860.9 & 56900.3 & $39.4 \pm 5.8$   & 1.8$\times10^{-8}$ & 4.9$\times10^{-4}$ & $9.1 \pm 1.4$ \\
        RBPLJ1751+0939  & OT 081 & 4FGL J1751.5+0938 & LSP & $134.8 \pm 3.1$  & 56901.9 & 56946.2 & $44.4 \pm 11.0$  & 2.4$\times10^{-5}$ & 6.3$\times10^{-2}$ & $3.0 \pm 0.8$ \\
        RBPLJ1751+0939  & OT 081 & 4FGL J1751.5+0938 & LSP & $220.1 \pm 1.9$  & 57206.9 & 57245.3 & $38.4 \pm 12.8$  & 4.4$\times10^{-7}$ & 1.6$\times10^{-2}$ & $5.7 \pm 2.0$ \\
        RBPLJ1754+3212  & ATO J268.5491+32.2064 & 4FGL J1754.2+3212 & ISP & $165.9 \pm 3.8$  & 57152.0 & 57219.9 & $67.9 \pm 11.9$  & 2.6$\times10^{-4}$ & 3.1$\times10^{-2}$ & $2.4 \pm 0.5$ \\
        RBPLJ1754+3212  & ATO J268.5491+32.2064 & 4FGL J1754.2+3212 & ISP & $99.5 \pm 4.0$   & 57235.4 & 57285.8 & $50.4 \pm 7.6$   & 2.9$\times10^{-6}$ & 3.1$\times10^{-2}$ & $2.0 \pm 0.3$ \\
        RBPLJ1754+3212  & ATO J268.5491+32.2064 & 4FGL J1754.2+3212 & ISP & $119.8 \pm 10.8$ & 57599.9 & 57621.9 & $22.0 \pm 2.9$   & 1.0$\times10^{-12}$ & 6.1$\times10^{-5}$ & $5.5 \pm 1.0$ \\
        RBPLJ1800+7828  & S5 1803+78 & 4FGL J1800.6+7828 & LSP & $111.1 \pm 2.4$  & 56513.9 & 56595.7 & $81.9 \pm 32.2$  & 5.2$\times10^{-4}$ & 9.8$\times10^{-4}$ & $1.4 \pm 0.6$ \\
        RBPLJ1800+7828  & S5 1803+78 & 4FGL J1800.6+7828 & LSP & $198.5 \pm 1.7$  & 56861.4 & 56898.8 & $37.4 \pm 6.5$   & 2.1$\times10^{-5}$ & 7.8$\times10^{-3}$ & $5.3 \pm 1.0$ \\
        RBPLJ1800+7828  & S5 1803+78 & 4FGL J1800.6+7828 & LSP & $156.8 \pm 3.4$  & 57225.5 & 57292.3 & $66.9 \pm 8.5$   & 1.5$\times10^{-4}$ & 1.6$\times10^{-2}$ & $2.4 \pm 0.3$ \\
        RBPLJ1800+7828  & S5 1803+78 & 4FGL J1800.6+7828 & LSP & $194.8 \pm 3.9$  & 57602.3 & 57620.3 & $18.0 \pm 7.3$   & 2.5$\times10^{-6}$ & 2.4$\times10^{-4}$ & $10.8 \pm 4.6$ \\
        RBPLJ1806+6949  & S4 1807+69 & 4FGL J1806.8+6949 & ISP & $174.0 \pm 4.5$  & 56789.6 & 56900.9 & $111.3 \pm 30.9$ & 4.0$\times10^{-8}$ & 9.8$\times10^{-4}$ & $1.6 \pm 0.5$ \\
        RBPLJ1838+4802  & ATO J279.7048+48.0428 & 4FGL J1838.8+4802 & HSP & $159.8 \pm 5.7$  & 57177.5 & 57274.9 & $97.4 \pm 20.1$  & 4.7$\times10^{-5}$ & 9.8$\times10^{-4}$ & $1.6 \pm 0.4$ \\
        RBPLJ2022+7611  & S5 2023+76 & 4FGL J2022.5+7612 & ISP & $136.8 \pm 2.3$  & 56892.9 & 56951.8 & $58.9 \pm 5.0$   & 1.5$\times10^{-3}$ & 6.3$\times10^{-2}$ & $2.3 \pm 0.2$ \\
        RBPLJ2149+0322  & MITG J2149+0323 & 4FGL J2149.6+0323 & ISP & $94.0 \pm 3.3$   & 56535.0 & 56592.3 & $57.3 \pm 8.5$   & 5.0$\times10^{-4}$ & 6.3$\times10^{-2}$ & $1.6 \pm 0.3$ \\
        RBPLJ2202+4216  & BL Lacertae & 4FGL J2202.7+4216 & LSP & $104.3 \pm 3.2$  & 57590.5 & 57604.9 & $14.4 \pm 4.4$   & 1.8$\times10^{-6}$ & 3.9$\times10^{-3}$ & $7.2 \pm 2.3$ \\
        RBPLJ2202+4216  & BL Lacertae & 4FGL J2202.7+4216 & LSP & $130.7 \pm 4.4$  & 57619.0 & 57637.0 & $18.1 \pm 1.5$   & 1.1$\times10^{-11}$ & 6.1$\times10^{-5}$ & $7.2 \pm 0.8$ \\
        RBPLJ2232+1143  & CTA 102 & 4FGL J2232.6+1143 & LSP & $314.9 \pm 12.2$ & 56500.1 & 56540.0 & $40.0 \pm 13.0$  & 4.5$\times10^{-4}$ & 3.9$\times10^{-3}$ & $7.9 \pm 2.8$ \\
        RBPLJ2232+1143  & CTA 102 & 4FGL J2232.6+1143 & LSP & $211.2 \pm 4.8$  & 56563.9 & 56609.8 & $45.8 \pm 10.0$  & 1.3$\times10^{-9}$ & 4.9$\times10^{-4}$ & $4.6 \pm 1.1$ \\
        RBPLJ2232+1143  & CTA 102 & 4FGL J2232.6+1143 & LSP & $188.3 \pm 3.6$  & 57205.6 & 57272.0 & $66.4 \pm 17.9$  & 6.9$\times10^{-5}$ & 6.3$\times10^{-2}$ & $2.8 \pm 0.8$ \\
        RBPLJ2232+1143  & CTA 102 & 4FGL J2232.6+1143 & LSP & $225.6 \pm 5.5$  & 57272.0 & 57336.3 & $64.4 \pm 13.5$  & 3.6$\times10^{-6}$ & 3.1$\times10^{-2}$ & $3.5 \pm 0.8$ \\
        RBPLJ2232+1143  & CTA 102 & 4FGL J2232.6+1143 & LSP & $121.0 \pm 4.5$  & 57598.5 & 57612.5 & $14.0 \pm 3.1$   & 2.7$\times10^{-13}$ & 2.0$\times10^{-3}$ & $8.6 \pm 2.1$ \\
        RBPLJ2232+1143  & CTA 102 & 4FGL J2232.6+1143 & LSP & $180.5 \pm 2.1$  & 57627.5 & 57637.0 & $9.5 \pm 2.2$    & 6.5$\times10^{-8}$ & 7.8$\times10^{-3}$ & $19.0 \pm 4.5$ \\
        RBPLJ2253+1608  & 3C 454.3 & 4FGL J2253.9+1609 & LSP & $178.8 \pm 5.3$  & 56494.6 & 56540.1 & $45.5 \pm 15.2$  & 6.3$\times10^{-10}$ & 2.4$\times10^{-4}$ & $3.9 \pm 1.4$ \\
        RBPLJ2253+1608  & 3C 454.3 & 4FGL J2253.9+1609 & LSP & $180.3 \pm 2.3$  & 56864.1 & 56885.5 & $21.5 \pm 2.4$   & 5.8$\times10^{-9}$ & 9.8$\times10^{-4}$ & $8.4 \pm 1.0$ \\
        RBPLJ2253+1608  & 3C 454.3 & 4FGL J2253.9+1609 & LSP & $136.9 \pm 2.9$  & 57274.0 & 57311.4 & $37.5 \pm 7.6$   & 2.4$\times10^{-2}$ & 6.3$\times10^{-2}$ & $3.7 \pm 0.8$ \\
        RBPLJ2253+1608  & 3C 454.3 & 4FGL J2253.9+1609 & LSP & $167.9 \pm 4.2$  & 57605.5 & 57612.6 & $7.0 \pm 2.1$    & 1.5$\times10^{-2}$ & 3.1$\times10^{-2}$ & $23.9 \pm 7.6$ \\
        \end{longtable}
      \end{minipage}
    }
  }
\end{center}

\endgroup

\FloatBarrier

\typeout{get arXiv to do 4 passes: Label(s) may have changed. Rerun}

\end{document}